\documentclass[twocolumn,showpacs,preprintnumbers,amsmath,amssymb,prb,floatfix]{revtex4}

\usepackage{graphicx}
\usepackage{dcolumn}
\usepackage{bm}
\usepackage{epstopdf} 
 
\begin{document}
\bibliographystyle{prsty}    

\title{Supercell studies of the Fermi surface changes in the electron-doped superconductor LaFeAsO$_{1-x}$F$_x$}
\author{P. Larson}
\author{S. Satpathy}%
\affiliation{%
Department of Physics, University of Missouri, Columbia, MO 65211  USA}%
\date{\today}

\begin{abstract}
We study the changes in the Fermi surface with electron doping in the LaFeAsO$_{1-x}$F$_x$ superconductors with density-functional supercell calculations
using the linearized augmented planewave (LAPW) method. The supercell calculations with explicit F substitution are compared with those obtained from the virtual crystal approximation (VCA) and from a simple rigid band shift. 
We find significant differences between the supercell results and those obtained from the rigid-band shift with electron doping, although quite remarkably the supercell results are in good agreement with the virtual crystal approximation (VCA) where the nuclear charges of the O atoms are slightly increased to mimic the addition of the extra electrons.
With electron doping, the two cylindrical hole pockets along $\Gamma-Z$ shrink in size, and the third hole pocket around $Z$ disappears for an electron doping concentration in excess of about 7-8\%, while the two elliptical electron cylinders along $M-A$ expand in size. The spin-orbit coupling does not affect the Fermi surface much except to somewhat reduce the size of the third hole pocket in the undoped case. We find that with the addition of the electrons the antiferromagnetic state becomes energetically less stable as compared to the nonmagnetic state, indicating that the electron doping may provide an extra degree of stability to the formation of the superconducting ground state.

\end{abstract}


\maketitle

\section{Introduction}

There has been an intense interest to understand the superconductivity of the recently discovered LaFeAsO.\cite{kamihara,wen,cruz,dong,haule,singh,yin,ma,xu,cao,yildirim,nekrasov,li,mazin,mazin2,kuroki,zhang,anisimov,opahle} Experiments have found values of the Curie temperature (T$_C$) as large as 26 K for electron doping of LaFeAsO$_{1-x}$F$_x$, 0.04 $\le x \le$ 0.12\cite{kamihara,wen}. Similar values of T$_C$ are found for hole doping of La with Sr but not with Ca\cite{wen,kamihara}. Neutron scattering\cite{cruz} and optical measurements\cite{dong} find an antiferromagnetic (AFM) ground state which has been confirmed by previous electronic structure calculations.\cite{haule,singh,xu,cao,ma,yin,yildirim,nekrasov,mazin} The nature of the superconductivity has not been understood, though evidence suggests its unconventional character.\cite{boeri,kuroki,mazin,mazin2,xu,cao,zhang}

The understanding of the normal-state electronic structure is important and serves as the foundation for understanding the superconductivity. One important question is what happens to the electronic structure when the extra electrons are added to the system via the fluorine dopants. A number of band structure studies have been performed to date to address these questions; however, most of them use either the simple rigid-band picture of shifting the Fermi energy in the band structure of the undoped system or the virtual crystal approximation.\cite{dong,haule,singh,yin,ma,xu,cao,yildirim,nekrasov,li,mazin,mazin2,kuroki,zhang} While these methods are expected to describe the rough picture,  the actual positions of the dopants could make significant differences to the band structure as compared to the rigid-band shift or to the VCA band structure, which is well known from the work on other systems.\cite{dargam,hass}
 In this work, we investigate the band structure using full supercell calculations
 and study the changes in the Fermi surface and the energetics with electron doping, with the fluorine substitution of the oxygen sites.
 


\section{Method of Calculation}

\begin{figure}[h!]
\centering
\includegraphics[width=7cm]{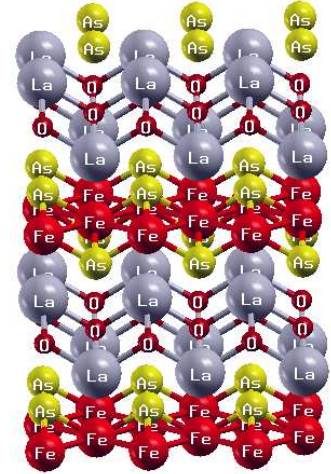}
\caption{(Color online) Crystal structure for LaFeAsO.}
\label{figcrystal}
\end{figure}

LaFeAsO forms in the $P4/nmm$ structure\cite{zimmer} with (Fe$^{2+}$As$^{3-}$)$_2$ layers lying between (La$^{3+}$O$^{2-}$)$_2$ layers, each of the atoms forming a square sublattice. Half of the As atoms belonging to the FeAs layer occur above the center of the Fe squares and the other half below it in an alternating pattern.
 They belong to a class of materials\cite{lebegue,xu} formed by one layer of a rare-earth atom with oxygen and another layer with late transition metal with a pnictogen atom. Each Fe atom, lying at the middle of a layer as seen in Fig. \ref{figcrystal}, is coordinated with four As atoms in distorted tetrahedral bonds above and below; O also lies in a distorted tetrahedron of La atoms. The doping of La (with Sr) or O (with F) is not in the magnetic FeAs layer but changes the magnetic properties nonetheless. Experimental lattice parameters of $a$ = 4.035 \AA\ and $c$ = 8.739 \AA\ were used. The internal parameters were relaxed by total energy minimization, the results of which agreed with the values reported in the literature\cite{singh}, viz., $z_{La}$ = 0.142 and $z_{As}$ = 0.633.  

Electronic structure calculations were performed using the linearized augmented plane wave (LAPW) method as implemented in the WIEN2k\cite{wien2k} program. The unit cell contains two formula units and for studying the effects of the dopants we used two supercells, a 16-atom supercell (four formula units) formed by doubling the cell in the $x$ or $y$ direction and a 32-atom supercell (eight formula unit)  formed by doubling the cell in the $xy$ plane in each direction. These two supercells correspond, respectively, to 25\% and 12.5\% F doping when one O atom is replaced by F. Calculations were also performed with the virtual crystal approximation (VCA)\cite{nordheim} with the standard unit cell. These two methods were used to understand the effects of F doping on the O sites. In the VCA the nuclear and the electron charge of the O atoms are increased continuously to approximate the additional electrons introduced by the F dopants.  For example, a 5\% concentration of F would change the nuclear and electronic charge of the O atoms from 8.0 to 8.05. Since superconductivity is expected to arise in the nonmagnetic (NM) state, we have focused on the electronic structure in the NM state. 

\begin{figure}[h!]
\centering
\includegraphics[width=8.4cm]{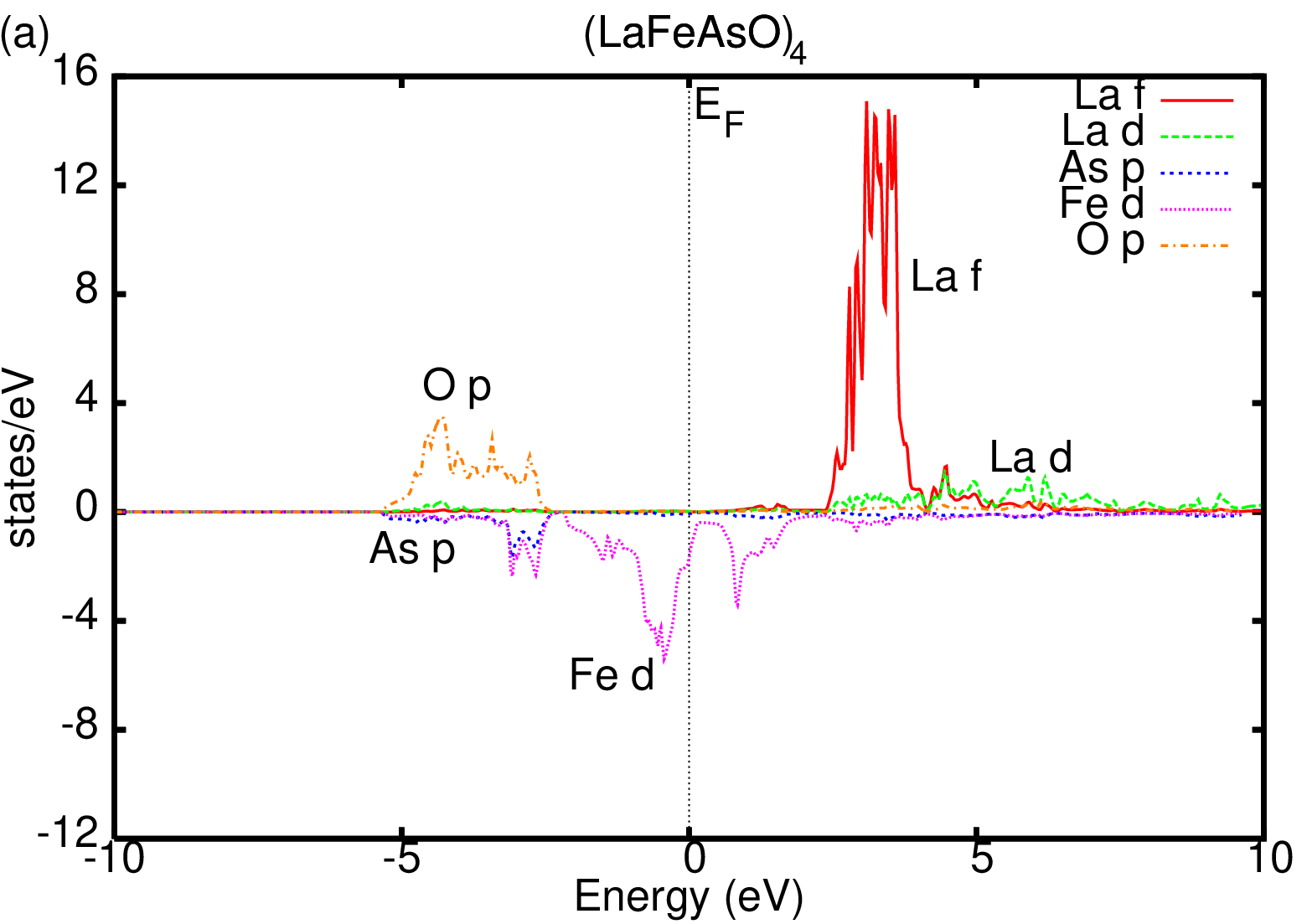}
\includegraphics[width=8.4cm]{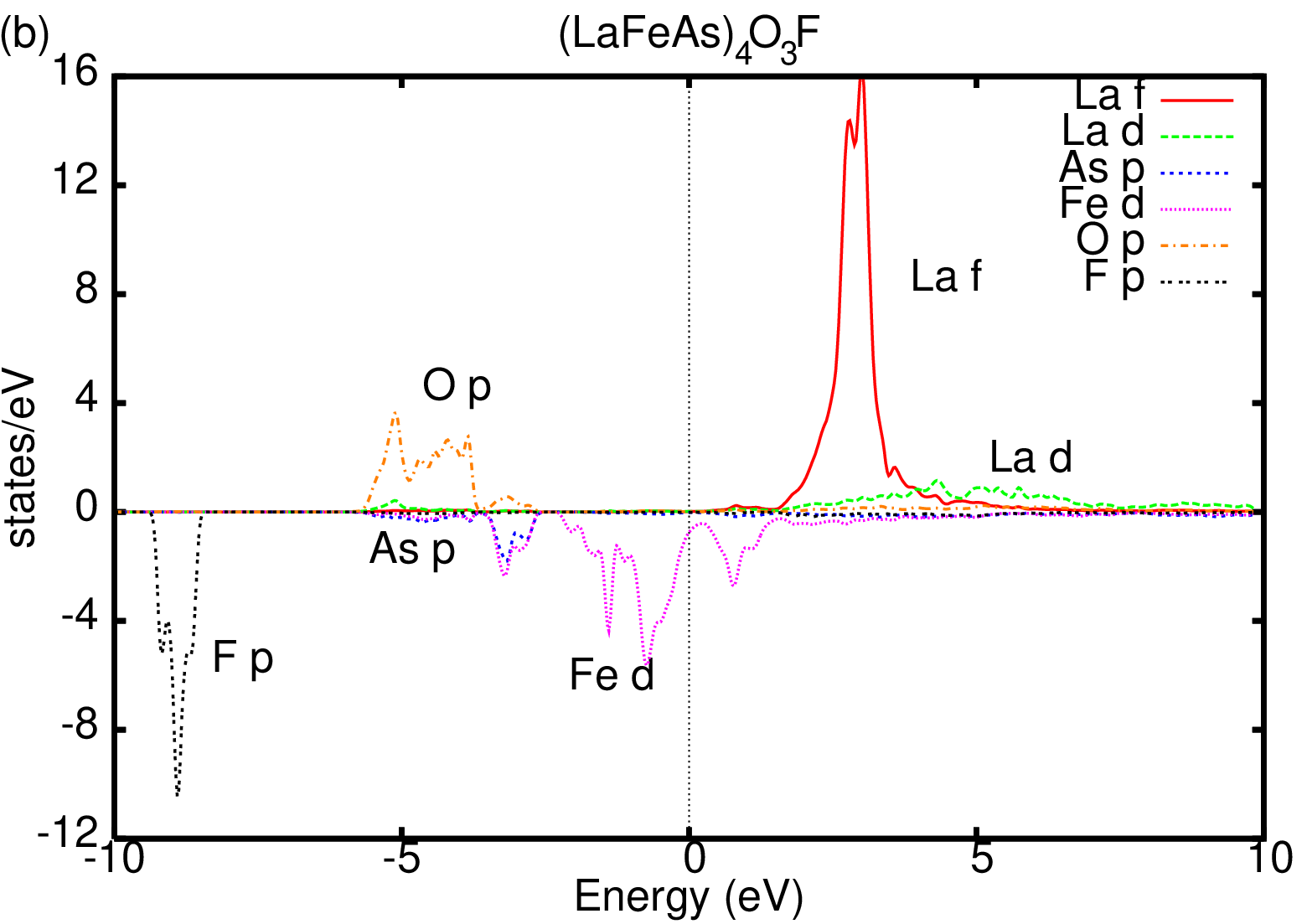}
\caption{(Color online) Density of states with or without fluorine doping calculated using the supercell LAPW method, indicating that the F states occur away from $E_F$.
}
\label{figdos1}
\end{figure}


In order to understand the effect of electron doping, we first discuss the results for the density of states obtained from the supercell calculation of F-doped LaFeAsO. The density of states (DOS) for LaFeAsO given in Fig. \ref{figdos1}a shows La $f$ and $d$ states lying above the Fermi level, while the O $p$ and As $p$ states occur below it. The O $s$ and As $s$ states lie well below, outside the range of the figure. The Fe $d$ states hybridize with the As $p$ states, though the size of the As sphere in the LAPW method leaves much of the As $p$ character outside the spheres, reducing its weight in the plot. This leaves the primary character of the bands observed in the calculated DOS near $E_F$ as Fe $d$. Strong Fe-Fe interactions cause the Fe $d$ states not to split apart into $t_{2g}$ and $e_g$ states. The positions of these states agree very well with those reported for the undoped LaFeAsO\cite{nekrasov,haule,yin,anisimov,singh,li} and LaFeAsP.\cite{lebegue} A full supercell calculation with 25\% F replacing O, shown in Fig. \ref{figdos1}b, finds that the F $p$ levels lie far below $E_F$ and act only to add electrons to the system, appearing to cause a rigid shift of the bands. As mentioned by previous authors\cite{ma}, although the total number of carriers increases, the electron doping shifts $E_F$ to a lower DOS, making it  hard to understand how the superconducting state can arise. However, while the DOS has a minimum at $E_F$, there is no evidence that the system is close to a metal-insulator transition.\cite{korotin} 

\begin{figure}[h!]
\centering
\includegraphics[width=6.0cm]{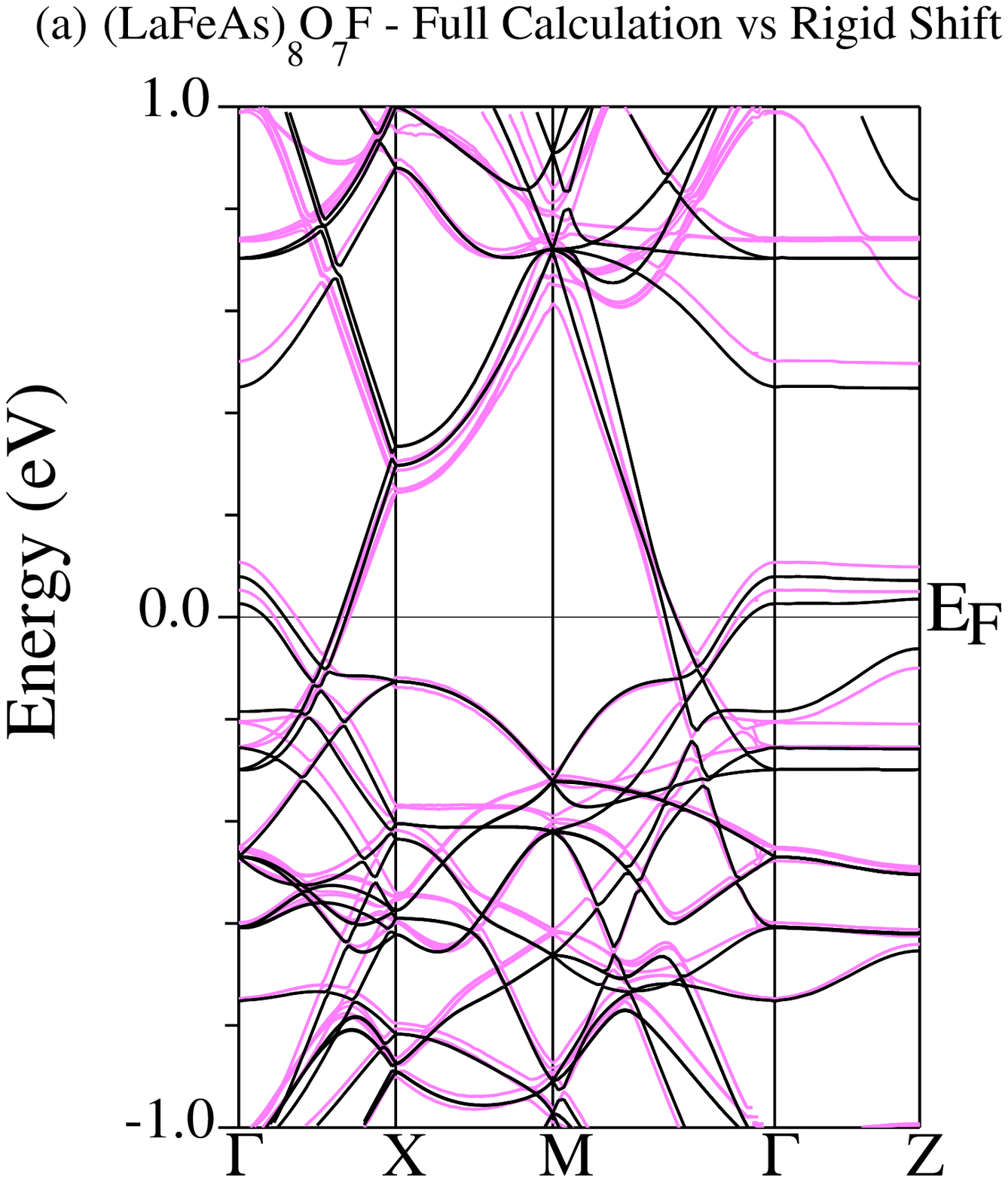}
\includegraphics[width=6.0cm]{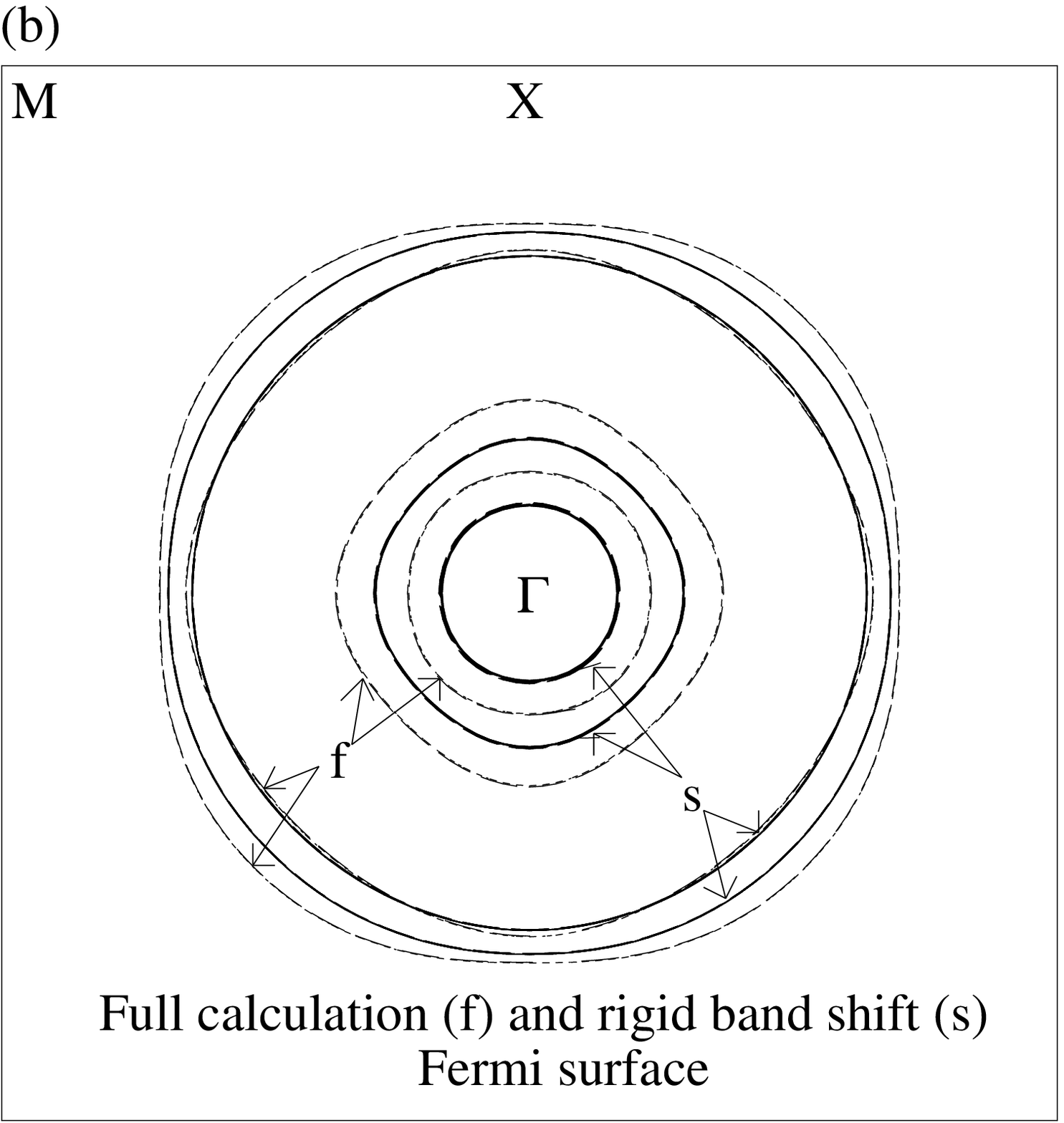}
\caption{(Color online) (a) Band structure for the 32-atom NM LaFeAsO unit cell shown for one F atom replacing O (La$_8$Fe$_8$As$_8$O$_7$F) in  violet and for the undoped material (La$_8$Fe$_8$As$_8$O$_8$) with rigid shift  in black and (b) the corresponding Fermi surfaces given on the $\Gamma-X-M$ plane. The symmetry points are for the supercell Brillouin zone, which has the same symmetry points as in the original unit cell but with half the magnitudes for the $k_x$ and $k_y$ components.}
\label{figbands2}
\end{figure}

From the calculated DOS (Fig. \ref{figdos1}), it might appear that the band structure for LaFeAsO is relatively unaffected by F doping, so that a rigid band shift of  $E_F$ to accommodate the added electrons might be good enough to describe the states at the Fermi energy. We find that while the overall shapes of the bands are about the same, there are enough differences in the states near $E_F$ to produce significant differences in the Fermi surface for the doped case. The band structure has been plotted in Fig. \ref{figbands2}a for the 32-atom supercell with one F atom on an O site and a calculation without F doping but with the bands rigidly shifted. In comparing the two cases, we have aligned the bands so that the energies of the deep oxygen core levels (O 1$s$ and 2$s$) remain the same, in view of the fact that
the deep core levels are very narrow in energy and they are not affected by the F substitution. 
Comparing the two sets of bands, the bands with F doping are sometimes above the shifted bands and sometimes below, so a better agreement is not possible simply by shifting the bands further. An important difference is the increased splitting of bands halfway between $\Gamma$ and $M$ at $E_F$. Previous calculations\cite{ma} have predicted that a rigid shift would lead to no separation between these two bands at $E_F$, but the supercell calculations show that these two bands remain apart.

Turning now to the Fermi surface, in the original Brillouin zone of the standard unit cell, the Fermi surface consists of two hole sheets around $\Gamma$ and two electron sheets around $M$. All sheets now occur around the $\Gamma$ point of the supercell Brillouin zone, since the original $M$ point gets folded to $\Gamma$. Most of the Fermi sheets in the full calculation have larger radii than that predicted from a rigid shift as the bands move further away from $\Gamma$ as seen from Fig. \ref{figbands2}b. Thus the rigid band shift does not describe very well the changes in the Fermi surface due to the doping.

\begin{figure}[h!]
\centering
\includegraphics[width=6.0cm]{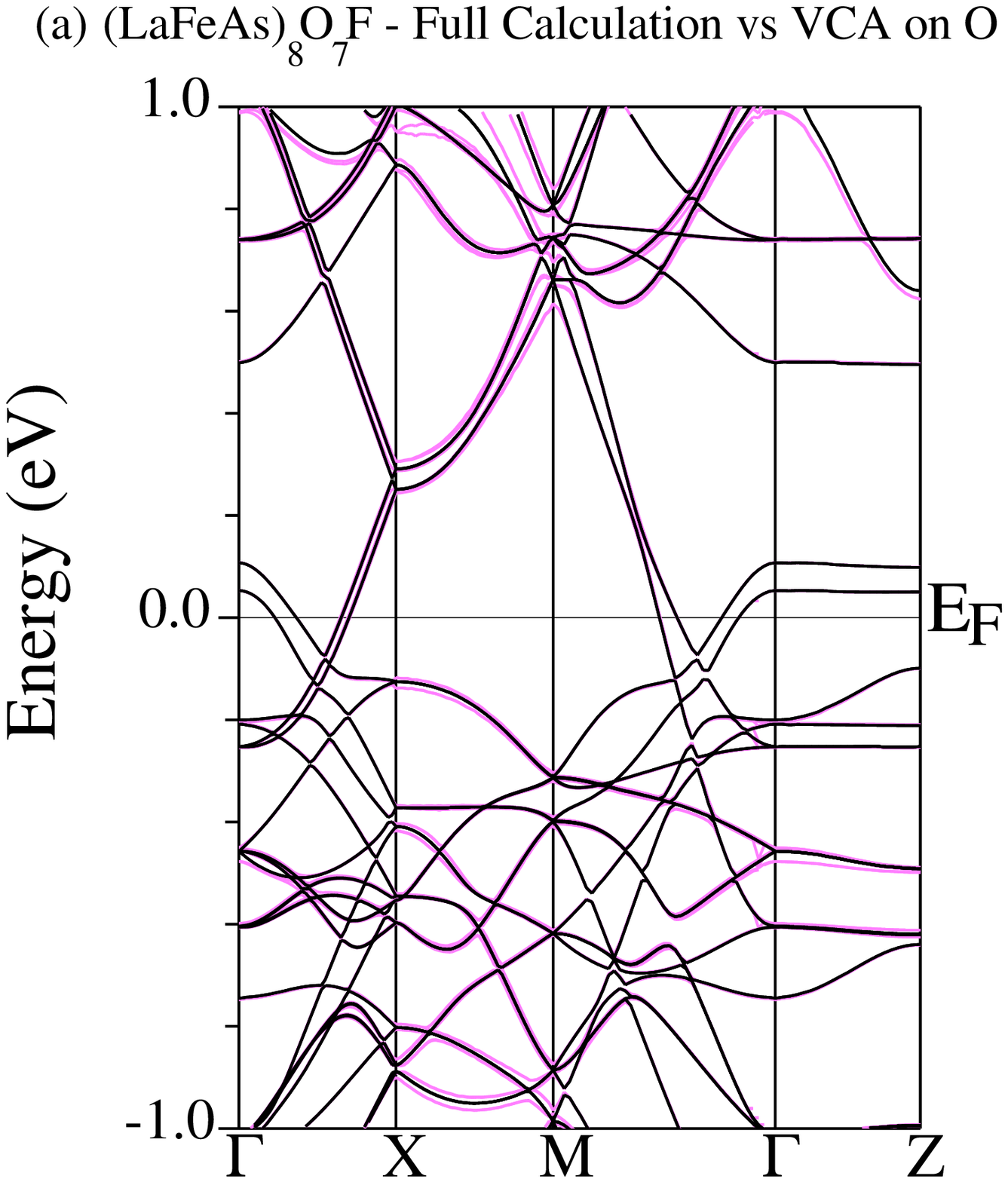}
\includegraphics[width=6.0cm]{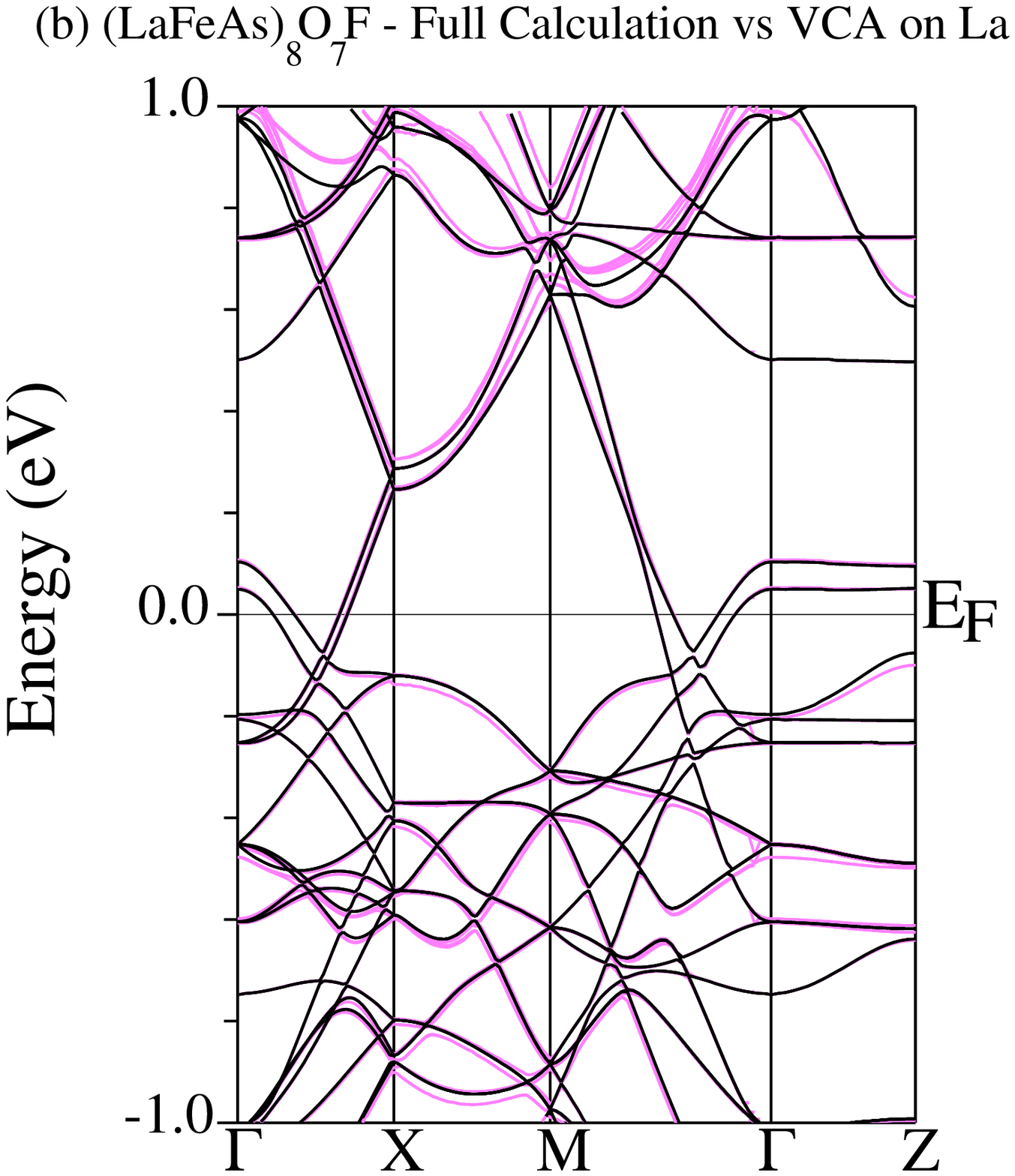}
\caption{(Color online) (a) Supercell band structure of fluorine doped LaFeAsO (La$_8$Fe$_8$As$_8$O$_7$F) (violet lines) compared with the equivalent VCA calculation (black lines) with changed O nuclear charge.
(b) Same as (a) except that the VCA calculation was done with changed La nuclear charge (black lines).  No difference is seen between the two sets of band structures near $E_F$.}
\label{figbands1}
\end{figure}

Our calculations of a rigid shift of the bands show significant changes in the Fermi surface compared to the full supercell calculation with the dopants included. In view of the fact that the states at $E_F$ are predominantly Fe $d$ and the F dopants are far from the FeAs layers, one might expect that the dopants could affect the band structure near $E_F$ in two ways: (a) by changing the Coulomb potential on different Fe sites by different amounts depending on their locations or (b) 
by introducing the extra electrons in the Fe layers which can then modify the on-site energies of different Fe orbitals differently because of their selective occupation of the various Fe($d$) orbitals. Quite interestingly, we find that there is a remarkable agreement between the VCA and the supercell results for states close to $E_F$ (Fig. \ref{figbands1}). In both cases, we have the same number of electrons in the FeAs layer and this agreement does not change even if we introduce the extra carriers in the VCA by changing the La nuclear charge instead of the O nuclear charge. This shows that the band structure is sensitive only to the electron concentration in the FeAs layer, so the Coulomb shift due to the relative position of the F dopants is lost by the dielectric screening due to the intermediate La and As layers. By the same token, the rigid band shift does not describe the band structure accurately because of the different concentration of the electrons implicit in the rigid band shift vs. the full calculation.

\begin{figure}[h!]
\centering
\includegraphics[width=5.5cm]{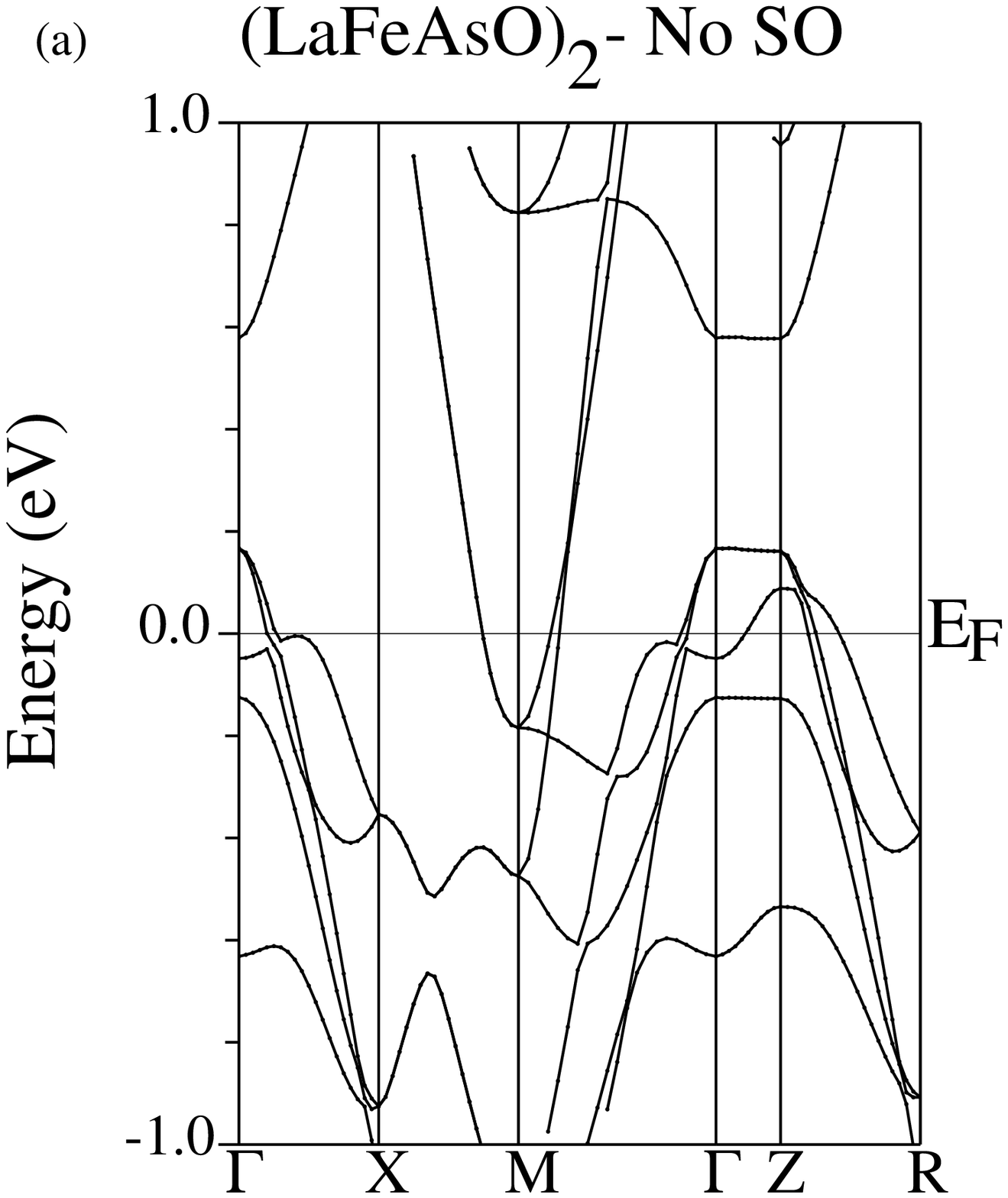}
\includegraphics[width=5.5cm]{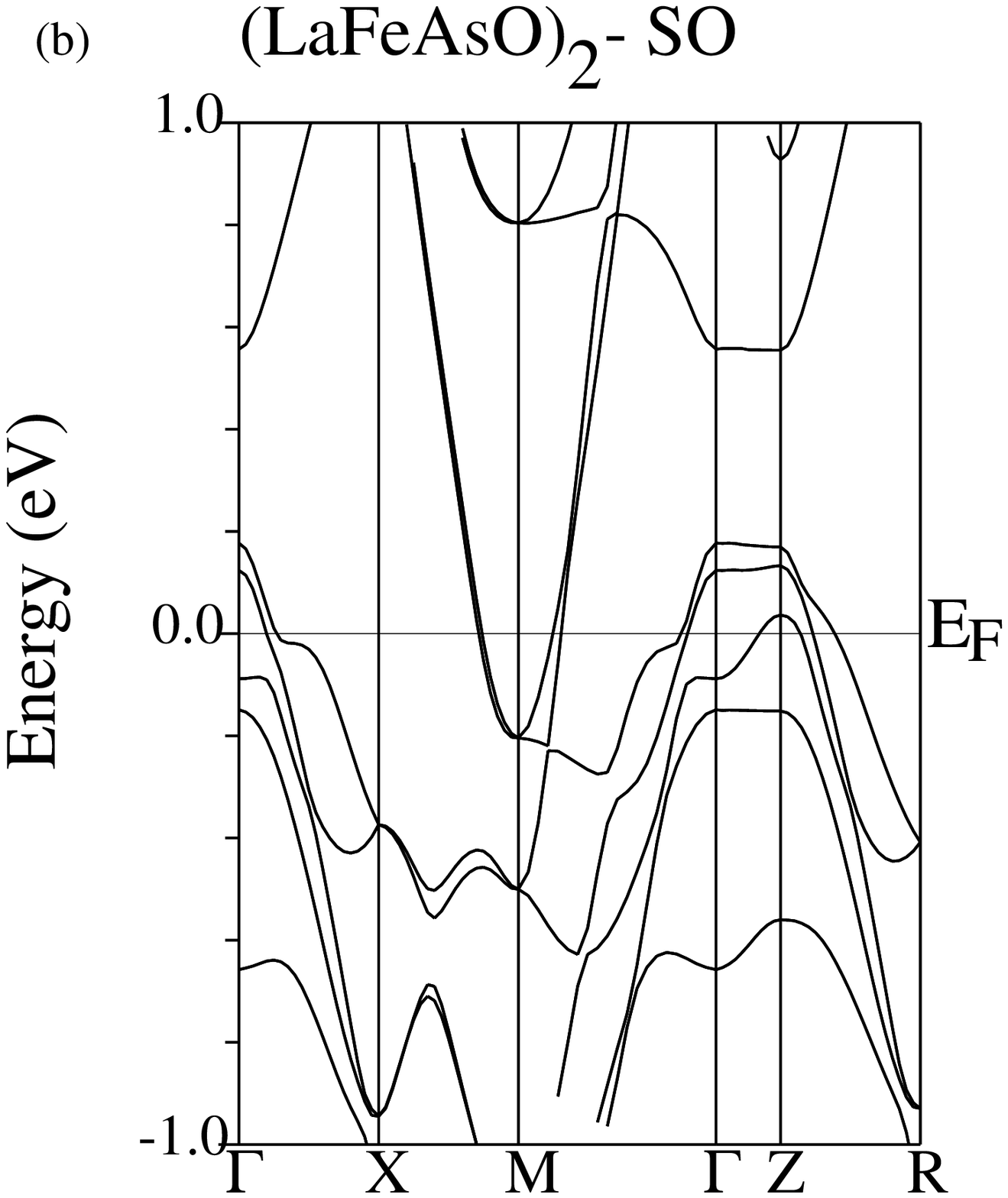}
\caption{Band structure of  (LaFeAsO)$_2$ with and without the spin-orbit interaction.}
\label{figbands2}
\end{figure}

While comparisons of VCA and rigid shifts of the bands are important to the Fermi surface, spin-orbit effects can also change the details of the Fermi surface. Since small changes to the Fermi surface can play an important role in superconductivity, spin-orbit effects cannot be ignored. Spin-orbit has not been investigated in LaFeAsO in any detail. Band structure calculations for the 8-atom unit cell given in Fig. \ref{figbands2}a,b show that spin-orbit lifts degeneracies for bands lying near $E_F$. One can see that the splitting is larger at $E_F$ along $\Gamma-X$ and $\Gamma-M$. Significant changes occur along $\Gamma-Z$ where the more dispersive Fe $d_{3z^2-1}$ band hybridizes with much less dispersive Fe $d_{xz}$ and $d_{yz}$ bands, separating the third hole pocket (with strong $d_{3z^2-1}$ character)  from the rest of the Fermi surface. The other bands at $E_F$ are relatively unchanged. 

\begin{figure}[h!]
\centering
\includegraphics[width=8.4cm]{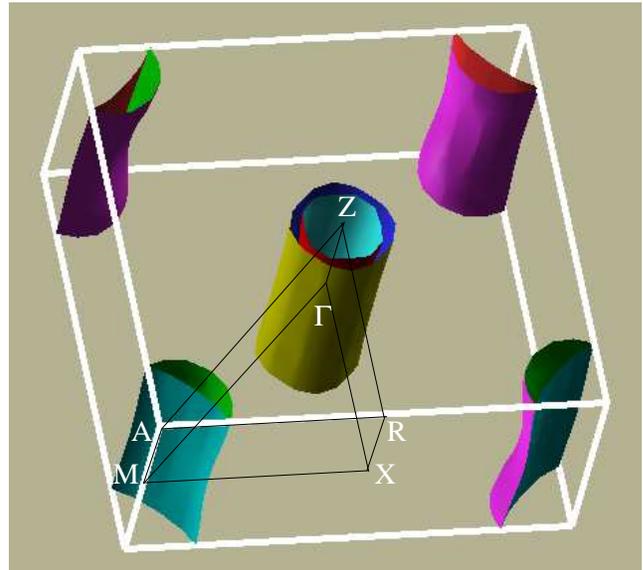}
\includegraphics[width=8.4cm]{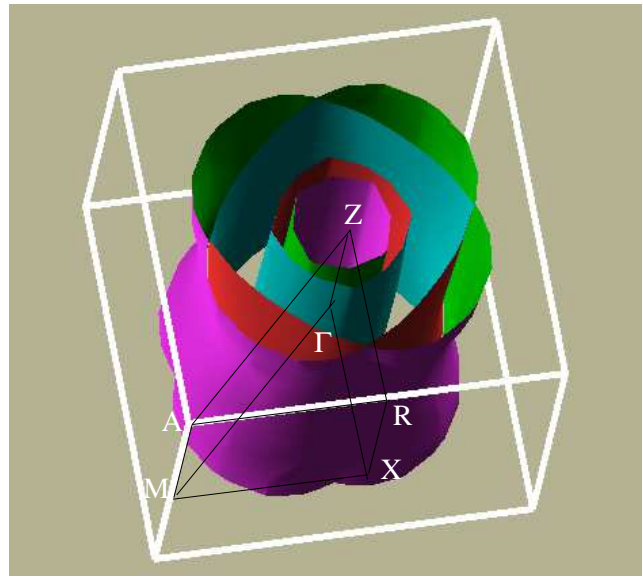}
\caption{(Color online) Fermi surface for (a) the 12.5\% fluorine dopant obtained from tha VCA shown in the standard Brillouin zone and (b) the full supercell calculation shown in the supercell Brillouin zone. }
\label{figfermi1}
\end{figure}

The calculated Fermi surface for the standard 8-atom unit cell corresponding to the undoped (LaFeAsO)$_2$ agrees well with previous calculations\cite{mazin,singh,ma,nekrasov,lebegue,xu}. We here show the Fermi surface calculated using the VCA for 12.5\% F concentration in Fig. \ref{figfermi1}a.  The Fermi surface consists of two cylindrical hole sheets lying along $\Gamma-Z$ and two cylindrical electron sheets lying along $M-A$. By doubling the unit cell in the both directions of the $xy$ plane to form the supercell, the Fermi surface undergoes band folding, as seen in Fig. \ref{figfermi1}b. This causes the elliptical electron pockets around $M$ to now surround the two cylindrical hole sheets at $\Gamma$. We note that the crystal symmetry of the supercell is the same as that as the smaller unit cell. Therefore, a point in the Brillouin zone, such as $M$ ($\frac{1}{2}$,$\frac{1}{2}$,0) is used in both figures, but corresponds to the fraction of the reciprocal lattice vectors in each case. Therefore the $M$ point in Fig. \ref{figfermi1}a is not the same $M$ point in Fig. \ref{figfermi1}b.  At lower concentrations, there exists a hole cap around the $Z$ point, as has been mentioned in previous calculations\cite{mazin,singh,ma,nekrasov,lebegue,xu}.

\begin{figure}[h!]
\centering
\includegraphics[width=6.0cm]{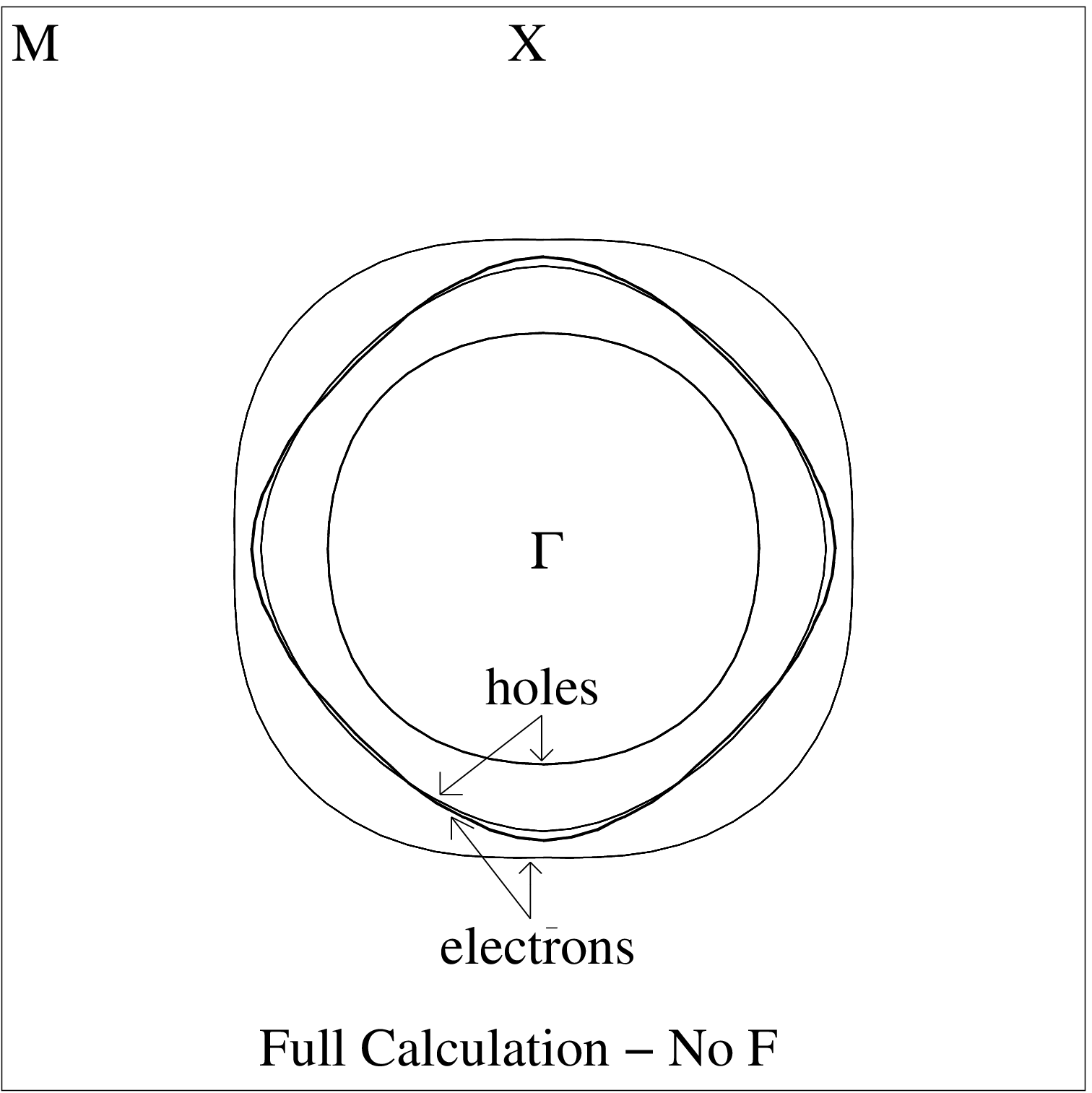}
\includegraphics[width=6.0cm]{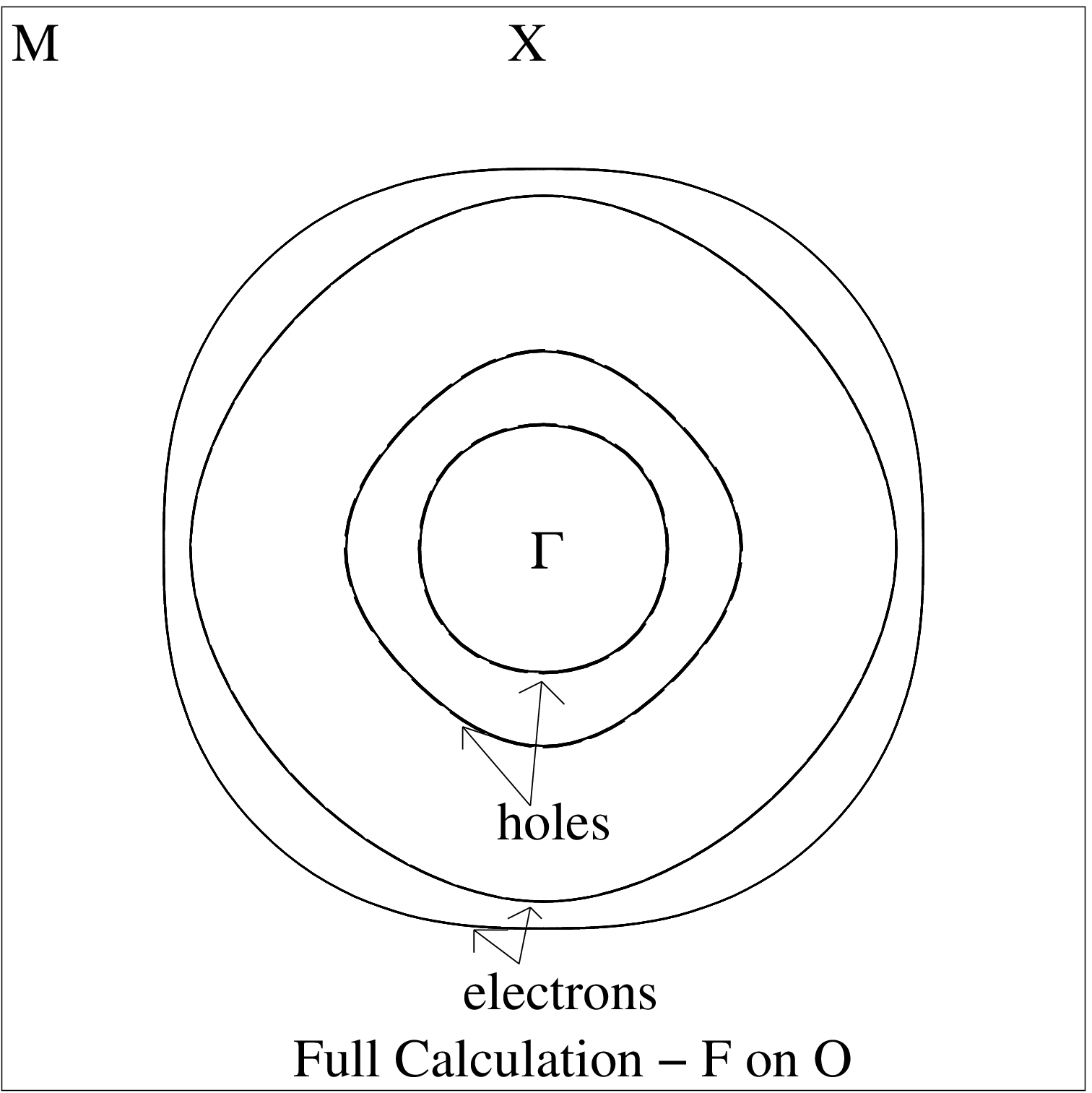}
\caption{Fermi surface shown in the $\Gamma-X-M$ plane of the supercell Brillouin zone for the 32-atom supercell of LaFeAsO with (a) no F doping and (b) with F replacing one of the 8 O sites.}

\label{figfermi3}
\end{figure}


Adding electrons to the FeAs plane via F doping increases the size of the elliptical electron pockets and  reduces  the size of the hole pockets. Full calculations performed in the 32-atom supercell with no F doping (Fig. \ref{figfermi3}a) shows the two elliptical electron pockets surrounding the two nearly circular hole pockets. The smaller electron pocket and the larger hole pocket nearly overlap. When we replace one F for an O atom (La$_8$Fe$_8$As$_8$O$_7$F) in Fig. \ref{figfermi3}b, the overall shape of the Fermi surface remains unchanged, but the electron pockets become larger while the hole pockets shrink. As we can see in the plot of the Fermi surface (Fig. \ref{figfermi1}), the hole pockets and electron pockets are narrower in the $\Gamma-X-M$ plane and become larger in the $Z-R-A$ plane. Addition of electrons reduces the differences between the sizes of the pockets in these two planes, so the Fermi surface looks more column-like in the $\Gamma-Z$ or $M-A$ direction, consistent with previous calculations.\cite{xu,lebegue,nekrasov,ma}  

\begin{table}
\caption{Radius $r_i$ of the hole pockets in \AA\  in the $\Gamma-M-X$ and $Z-A-R$ planes at different doping levels $x$. The full supercell calculations with F dopants agree with the VCA results for the cases where we have compared them ($x$=0, 0.125, and 0.25). The third hole pocket on the second plane   disappears beyond $x=0.07$.}

\begin{tabular}{c|cc|ccc}
\tableline
x&$\Gamma MX$ plane&&$ZAR$ plane&&\\
\hline
&$r_1$&$r_2$&$r_1$&$r_2$&$r_3$\\
\hline
0.00&0.101&0.085&0.168&0.094&0.060\\
0.07&0.105&0.084& 0.151 & 0.096 & 0.054\\
0.08&0.087&0.066& 0.093 & 0.076 & -\\
0.125&0.078&0.054&0.076&0.054&-\\
0.25&0.078&0.030&0.073&0.013&- \\
\hline
\end{tabular}
\label{tabl1}
\end{table}

\begin{table}
\caption{Major and minor axes and eccentricities (a, b, and $\epsilon$, respectively) of electron pockets at different doping levels $x$ obtained from the supercell calculations. }

\begin{tabular}{c|ccc|ccc}
\tableline
x&&$\Gamma MX$ plane&&&$ZAR$ plane&\\

\tableline
&a (\AA)&b (\AA)&$\epsilon$&a (\AA)&b (\AA)&$\epsilon$\\
\tableline
0.00&0.102&0.128&0.601&0.102&0.152&0.744 \\
0.125&0.104&0.127&0.571&0.104&0.147&0.707\\
0.25&0.122&0.148&0.571&0.122&0.172&0.707\\
\tableline
\end{tabular}
\label{tabl2}
\end{table}

The size of the hole and electron pockets were calculated using the VCA and the supercell calculation with F substitution on O sites and have been shown in Tables I and II. Since the Fermi surfaces obtained from the VCA and the full supercell calculations are substantially the same, only one number is given for each concentration. The size of the electron and hole pockets were calculated along the $\Gamma-M$ direction. 
The hole pockets as shown in Fig. \ref{figfermi1} are circular (or nearly so) lying along $\Gamma-Z$ and consist of hybridized Fe $d_{xz}$ and $d_{yz}$ states. The $d_{xz}$ and $d_{yz}$ orbitals are degenerate due to the point group symmetry of Fe. This is consistent with previous calculations\cite{yin,ma}. There exists a third band which forms the cap around the $Z$ point, mostly of $d_{3z^2-1}$ character. This third Fermi surface sheet disappears below $E_F$ with 7-8\% electron doping.  

The electron pockets are elliptical with significant nesting characteristics. Several proposed superconducting theories require understanding of the eccentricity of electron pockets which affects the Fermi surface nesting and in addition may be important for magnetic instabilities\cite{mazin}. We list in Table II the calculated eccentricity as a function of electron doping using the standard definition $\epsilon$ = $\sqrt{1-b^2/a^2}$, where $a$ and $b$ are, respectively, the major and the minor axes.
The two electron Fermi surface sheets surround the $M-A$ points in the standard unit cell (Fig. \ref{figfermi1}a). The electron cylinders arise out of  two bands, one of which is primarily of Fe $d_{x^2-y^2}$ character and the other, of mixed Fe $d_{xz}$ and $d_{yz}$ character. These two bands can be identified as those lying along M$\Gamma$ in Fig. \ref{figbands2}a, b crossing about 0.25 eV above $E_F$. The eccentricity of the ellipse arises due to different dispersion along different directions in the $xy$ plane.\cite{note}
With electron doping, the separation between these two bands decreases along $\Gamma-M$, reducing the eccentricity. However, unlike what was seen in a rigid shift of the bands\cite{ma}, the eccentricity never disappears or begins to increase with electron doping.

\begin{figure}[h!]
\centering
\includegraphics[width=8.4cm]{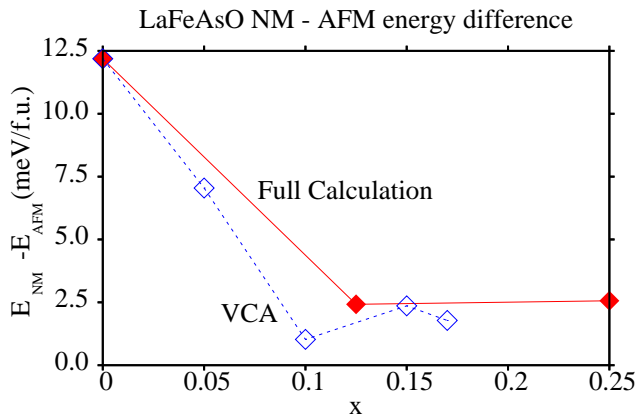}
\caption{Energy difference between the nonmagnetic and the antiferromagnetic state as a function of electron doping for  LaFeAsO$_{1-x}$F$_x$.
}
\label{figene}
\end{figure}


Conventional theories of the superconductivity describe the superconducting state to arise from the Fermi surface instability of the paramagnetic normal state, while density functional calculations show the 
ground state of the undoped material to be an antiferromagnetic metal. 
Therefore the question arises as to whether the electron doping destabilizes the AFM state in favor of a paramagnetic state thereby facilitating the formation of the superconducting state. To address this question, we have performed calculations of the total energy with and without electron doping in the supercell geometry and have shown these results in Fig. \ref{figene} along with the VCA results.
The results of the full supercell calculation and the VCA energies agree quite well, which is consistent with the excellent agreement between their two band structures (Fig. \ref{figbands1}). We find that even though the AFM state is stable for all dopant concentrations, the energy of the NM state is significantly reduced as compared to that of the AFM state.
These results suggest that the the electron doping might serve to destabilize the AFM state in favor of the nonmagnetic state thereby facilitating superconductivity.


\section{Summary}

In summary, from density-functional supercell calculations we have studied the changes in the Fermi surface of  LaFeAsO as a function of electron doping. Important differences in the Fermi surface were found from results obtained with the simple rigid-band shift, while the virtual crystal approximation yielded reasonable results. Finally, our total energy results suggest that electron doping might provide an extra degree of stability to the superconducting state by making the AFM normal state less favorable. 


\acknowledgments{This work was supported by the U.S. Department of Energy under Grant No. DE-FG02-00ER45818.}


\begin{thebibliography}{99}

\bibitem{wen} H.-H. Wen, G. Mu, L. Fang, H. Yang, and X. Zhu, Europhys. Lett. {\bf 82}, 17009 (2008).

\bibitem{kamihara} J. Amer. Chem. Soc. {\bf 130}, 3296 (2008).

\bibitem{cruz} C. de la Cruz, Q. Huang, J.W. Lynn, J. Li, W. Ratcliff II, J.L. Zaretsky, H.A. Mook, G.F. Chen, J.L. Luo, N.L. Wang, and P. Dai, arXiv:0804.0795.

\bibitem{dong} J. Dong, H.J. Zhang, G. Xu, Z. Li, W.Z. Hu, D. Wu, G.F. Chen, X. Dai, J.L. Luo, Z. Fang, and N.L. Wang, arXiv:0803.3426.

\bibitem{haule} K. Haule, J.H. Shim, and G. Kotliar, Phys. Rev. Lett. {\bf 100}, 226402 (2008).

\bibitem{singh} D.J. Singh and M.-H. Du, arXiv:0803:0429.

\bibitem{yin} Z.P. Yin, S. Leb\'{e}gue, M.J. Han, B. Neal, S.Y. Savrasov, and W.E. Pickett, arXiv:0804.3355.

\bibitem{ma} F. Ma and Z.-Y. Lu, arXiv:0803.3286

\bibitem{xu} G. Xu, W. Ming, Y. Yao, X. Dai, S.-C. Zhang, and Z. Fang, arXiv:0803.1282.

\bibitem{cao} C. Cao, P.J. Hirschfeld, and H.-P. Cheng, arXiv:0803.3236.

\bibitem{yildirim} T. Yildirim, arXiv:0804.2252.

\bibitem{nekrasov} I.A. Nekrasov, Z.V. Pchellkina, and M.V. Sadovskii, arXiv:0804.1239.

\bibitem{mazin} I.I. Mazin, M.D. Johannes, L. Boeri, K. Koepernik, and D.J. Singh, arXiv:0806.1869.

\bibitem{kuroki} K. Kuroki, S. Onari, R. Arita, H. Usui, Y. Tanaka, H. Kontani, and H. Aoki, arXiv:0803.3325.

\bibitem{zhang} H.-J. Zhang, G. Xu, X. Dai, and Z. Fang, arXiv:0803.4487.

\bibitem{li} H. Li, J. Li, S. Zhang, W. Chu, D. Chen, and Z. Wu, arXiv:0807.3153.

\bibitem{mazin2} I.I. Mazin, D.J. Singh, M.D. Johannes, and M.H. Du, Phys. Rev. Lett {\bf 101}, 057003 (2008).

\bibitem{anisimov} V.I. Anisimov, Dm. M. Korotin, S.V. Stretlsov, A.V. Kozhevnikov, J. Kune\~{s}, A.O. Shorikov, and M.A. Korotin, arXiv:0807.0547. 

\bibitem{opahle} I. Opahle, H.C. Kandpal, Y. Zhang, C. Gros, and R. Valent\`{i}, arXiv:0808.0834. 

\bibitem{boeri} L. Boeri, O.V. Dolgov, and A.A. Golubov, arXiv:0803.2703; L. Shan, W. Wang, X. Zhu, G. Mu, L. Fang, and H.-H. Wen, arXiv:0803.2405.

\bibitem{dargam} T.G. Dargam, R.B. Capaz, and B. Koiller, Braz. Jour. Phys. {\bf 27/A}, 299 (1997).

\bibitem{hass} K.C. Hass and R.J. Baird, Phys. Rev. B {\bf 38}, 3591 (1988).

\bibitem{zimmer} B.I. Zimmer, W. Jeitschko, J.H. Albering, R. Glaum, M. Reehuis, J. Alloys Compd. {\bf 229}, 238 (1995).

\bibitem{lebegue} S. Lebegue, Phys. Rev. B {\bf 75}, 035110 (2007).

\bibitem{wien2k} P. Blaha, K. Schwarz, G.K.H. Madsen, D. Kvasnicka, and J. Luitz in {\it WIEN2k, An Augmented Plane Wave Plus Local Orbitals Program for Calculating Crystal Properties}, edited by K. Schwarz (Technische Universit\"{a}t Wien, Austria, 2001).

\bibitem{nordheim} L. Nordheim, Ann. Phys. (Leipzig) {\bf 9}, 607 (1931).

\bibitem{korotin} M.A. Korotin, S.V. Streltsov, A.O. Shorikov, and V.I. Anisimov, arXiv:0805.3453; A.O. Shorikov, M.A. Korotin, S.V. Streltsov, D.M. Korotin, V.I. Anisimov, and S.L. Skornyakov, arXiv:0804.3283.

\bibitem {note} Note that while we have used the standard definition of ellipticity, sometimes an alternate definition is used in the literature. 

\end{thebibliography}
\end{document}